\documentclass[twocolumn,showpacs,preprintnumbers]{revtex4}
\usepackage{graphicx}% Include figure files
\usepackage{dcolumn}% Align table columns on decimal point
\usepackage{bm}% bold math

\begin{document}

%%%%%%%%%%%%%%%%%%%%%%%%%%%%%%%%%%%%%%%%%%%%%%%%%%%%%%%%%%%%%%%%%%%%%%%%%%%%
%%%%%%%%%%%%%%%%%%%%%%%%%%%%%%%%%%%%%%%%%%%%%%%%%%%%%%%%%%%%%%%%%%%%%%%%%%%%

\title{Signatures of Electronic Nematic Phase at Isotropic-Nematic Phase Transition}
\author{Hae-Young Kee$^1$, Eugene H. Kim$^{1,2}$, and Chung-Hou Chung$^1$  }
\affiliation{$^1$ Department of Physics, University of Toronto, 
Toronto, Ontario, Canada M5S 1A7  \\
$^2$ Department of Physics and Astronomy, McMaster University, 
Hamilton, Ontario, Canada L8S 4M1 }
\date{\today}

\begin{abstract}
The electronic nematic phase occurs when the point-group 
symmetry of the lattice structure is broken, due to 
electron-electron interactions.  
We study a model for the nematic phase 
on a square lattice with emphasis on the phase transition 
between isotropic and nematic phases within mean field theory.  
We find the transition to be first order,
with dramatic changes in the Fermi surface 
topology accompanying the transition.  
Furthermore, we study 
the conductivity tensor and Hall constant as probes of the nematic 
phase and its transition.  The relevance of our findings to Hall 
resistivity experiments in the high-$T_c$ cuprates is discussed.

%Recently, electronic nematic phases --- anisotropic phases with broken 
%rotational symmetry, but with translational symmetry in tact
%--- have been argued to play a role in various strongly correlated 
%electron systems.  

%We study the phase 
%transition of the isotropic Fermi liquid to the nematic 
%Fermi liquid by solving the self-consistent mean field 
%equations of the Hamiltonian with quadrupolar 
%density-density interaction on a square lattice.  The 
%nematic order develops in a region of finite doping 
%concentration as a first order transition, leading 
%to a dramatic change in the Fermi surface topology.
%We investigate the effect of nematic order on the 
%conductivity tensor.  We find that the Hall constant 
%in the weak field limit is discontinuous at the phase 
%transition between the isotropic and nematic Fermi 
%liquids.  The relevance of our findings to Hall 
%resistivity experiments in the high-$T_c$ cuprates is 
%discussed.
\end{abstract}         

\maketitle

\def\be{\begin{equation}}
\def\ee{\end{equation}}
\def\bea{\begin{eqnarray}}
\def\eea{\end{eqnarray}}
 
%%%%%%%%%%%%%%%%%%%%%%%%%%%%%%%%%%%%%%%%%%%%%%%%%%%%%%%%%%%%%%%%%%%%%%%%%%%%
%%%%%%%%%%%%%%%%%%%%%%%%%%%%%%%%%%%%%%%%%%%%%%%%%%%%%%%%%%%%%%%%%%%%%%%%%%%%

\section{Introduction}

Recently, novel spatially inhomogeneous and/or anisotropic 
phases have been found to occur in various condensed matter 
systems.  Similar to classical liquid crystals,\cite{de_gennes} 
these quantum phases can be classified according to their 
broken symmetries.  
One such phase is the smectic phase, often referred to as a 
stripe phase.  This phase can be viewed as an unidirectional 
charge density wave phase which spontaneously breaks 
translational symmetry along one direction.  Experimental
evidence for smectic phases has been found in several 
transition-metal oxide compounds.  In particular, a stripe 
crystal phase has been observed in manganese oxide compounds 
by electron diffraction.\cite{cheong}  Moreover, neutron 
scattering experiments have provided indirect evidence for 
stripe phases in the high-$T_c$ cuprate 
superconductors.\cite{tranquada}

Another such phase is the nematic phase --- an anisotropic phase 
with broken rotational symmetry, but with translational symmetry 
in tact.  This phase can be thought of as a smectic phase ``melted'' 
via the proliferation of topological defects.  
Nematic phases have been argued to be the source of the anisotropic
transport observed in quantum Hall systems in high Landau 
levels\cite{eisenstein,fradkin}, and to be relevant to the 
high-$T_c$ cuprate superconductors as well.\cite{kivelson}
However, an understanding of the transition from the smectic
to nematic phase due to the proliferation of the topological 
defects still remains to be achieved.

Recently, a complementary approach to studying the nematic 
phase was suggested.\cite{vadim,metzner}  In this approach, the 
nematic phase is obtained from an isotropic liquid by 
breaking rotational symmetry.  In the continuum limit, it 
was shown that the Goldstone mode resulting from breaking 
rotational symmetry leads to non-Fermi liquid behavior in 
the single-particle lifetime.\cite{vadim}  Exotic superconducting 
instabilities mediated by the collective mode in the nematic 
phase were also studied.\cite{kim}

% The evidence of the existence of an anisotropic collective state of matter
% has been reported in the ultra-clean two dimensional electron systems
% with high magnetic field.\cite{eisenstein}

%%%%%%%%%%%%%%%%%%%

%Though the two approaches give (nematic) phases having the 
%same symmetries, it is not clear whether they are actually 
%the same phase.
%%
%More importantly, however, it is not clear which approach
%is more appropriate to describe the phenomena occuring in
%physical systems.
%%
%Here, we pursue the latter approach.  Our goals are to
%better understand this model, as it is interesting in 
%its own right; as in Ref.~\onlinecite{kee,kivelson2}, 
%deduce consequences of nematic order to probe the nematic 
%state.
%%
%More importantly, however, we would like to suggest 
%experiments that could help us learn which approach is
%most relevant for describing physical systems.
%%
While the consequences of the nematic order 
in the recent complementary approach were
studied in Ref. ~\onlinecite{kee,kivelson2},
we still lack an understanding of the phase transition between
the  isotropic and nematic phases.
In this work, we study a model for the nematic phase on a
square lattice, with emphasis on the phase transitions
between the isotopic and nematic phases.  Furthermore, we 
study the conductivity tensor and the Hall constant as 
probes of nematic phase and its transition.

%%%%%%%%%%%%%%%%%%%%%%%%%%%%%%%%%%%%%%%%%%%%%%%%%%%%%%%%%%%%%%%%%%%%%%%%%%%%
%%%%%%%%%%%%%%%%%%%%%%%%%%%%%%%%%%%%%%%%%%%%%%%%%%%%%%%%%%%%%%%%%%%%%%%%%%%%

\section{Model Hamiltonian and Phase Transitions}

The Hamiltonian we consider is 
\begin{equation}
H = \sum_{\bf k} \xi_{\bf k} 
 c_{\bf k}^{\dagger} c^{\phantom \dagger}_{{\bf k}}
 + \sum_{\bf q} F_2({\bf q})
{\rm Tr} \left[ {\hat Q} ({\bf q}) {\hat Q} (-{\bf q}) \right].
\label{model}
\end{equation}
In Eq.~\ref{model}, $\xi_{\bf k}$ is the single-particle 
dispersion; $F_2({\bf q})$ is the inter-electron interaction 
strength, which can be written as
\[
F_2 ({\bf q}) = \frac{F_2}{1+\kappa q^2} \, ;
\]
${\hat Q}({\bf q})$ is the quadrupole density tensor
\[
{\hat Q} ({\bf q}) = \frac{1}{\sqrt{N}} \sum_{\bf k}
c_{{\bf k}+\frac{{\bf q}}{2} }^{\dagger}
\left (\matrix{\cos{k_x} - \cos{k_y} &  \sin{k_x}\sin{k_y} \cr
 \sin{k_x} \sin{k_y}   & \cos{k_y} - \cos{k_x} \cr} \right )
c^{\phantom \dagger}_{{\bf k}-\frac{{\bf q}}{2} } \, ,
\]
where $N$ is the number of lattice sites. 
In what follows, we take $\xi_{\bf k}$ to arise from a 
tight-binding model on a square lattice with nearest-neighbor 
($t$) and next-nearest neighbor ($t^{\prime}$) hopping integrals
\[
 \xi_{\bf k} = - 2t (\cos{k_x}+\cos{k_y} ) 
- 4 t^{\prime} \cos{k_x} \cos{k_y} - \mu \, ,
\]  
where $\mu$ is the chemical potential, and we have set the lattice
spacing to unity, $a = 1$.

Within a mean field approximation, we have the self-consistency
equations
\begin{eqnarray}
\langle Q_{xx} \rangle &=& \frac{1}{N} \sum_{\bf k} (\cos{k_x} -\cos{k_y}) 
   f(\xi_{\bf k}) \, , \nonumber\\
\langle Q_{xy} \rangle &=& \frac{1}{N} \sum_{\bf k} ( \sin{k_x} \sin{k_y}) 
   f(\xi_{\bf k}) \, ,  \label{nematic} \\
n &= & \frac{1}{N} \sum_{\bf k} f(\xi_{\bf k}) \, ,  \nonumber
\end{eqnarray}
where $f(\xi_{\bf k})$ is the Fermi function, and $n$ is the
electron density.  The expectation values $\langle Q_{xx} \rangle$ 
and $\langle Q_{xy} \rangle$ lead to a modification of the
single-particle dispersion.  Defining
\begin{eqnarray}
 \lambda = |F_2| \langle Q_{xx} \rangle/(2 t) \ \ {\rm and} \ \
 \lambda^{\prime} = |F_2| \langle Q_{xy} \rangle /(4 t^{\prime}) \, ,  
\end{eqnarray}
the single particle dispersion becomes
\begin{eqnarray}
 \xi_{\bf k} & =  & -2 t \left[ \left( 1 +  \lambda \right) \cos{k_x}
    + \left(1- \lambda \right)  \cos{k_y}  \right] \nonumber\\
  & - & 4 t^{\prime} (\cos{k_x} \cos{k_y} +  \lambda^{\prime}
 \sin{k_x} \sin{k_y})  -  \mu \, .
\label{band}
\end{eqnarray}
For given values of $\mu$, $F_2$, and $t^{\prime}$, 
we solved for $\lambda$, $\lambda^{\prime}$ and the
density $n=1-x$ ($x$ is the hole concentration)
using Eq.~\ref{nematic}.
%For given values of $\mu$ which determines the
%density, $n = 1-x$ ($x$ is the hole concentration),
%$F_2$, and $t^{\prime}$, we solved for $\lambda$, 
%$\lambda^{\prime}$, and $n$ using Eq.~\ref{nematic}.
For all values of the parameters we considered, we found 
$\lambda^{\prime} = 0$.  

\begin{figure}
\scalebox{.44}{\includegraphics{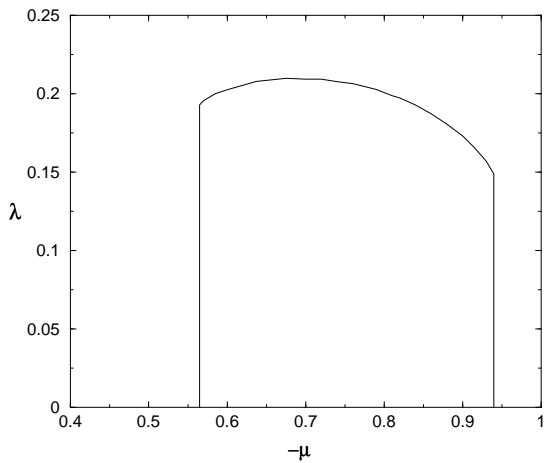} }
\caption{The nematic order $\lambda$ as a function of chemical potential, $\mu$
for $|F_2|/(2 t) =0.55$ and $t^{\prime} =-0.4 t$ ($2 t\equiv 1$) at $T=0$.
$\mu=-0.4$ corresponds to the half filling, $x=0$ }
\label{fig:lambdaplot}
\end{figure}

In Fig.~\ref{fig:lambdaplot}, we show the behavior of the 
nematic order $\lambda$ as a function of chemical potential,
$\mu$ for $F_2/(2t) = -0.55$, $t^{\prime} =-0.4 t$,
and $2 t \equiv 1$ at $T=0$.  
An important feature of 
Fig.~\ref{fig:lambdaplot} is the sharp onset and offset of 
nematic order at critical values of $\mu$:
$\mu_c=-0.57$ and $-0.94$.  
The free energy
near the transition ---  $-0.61 \le \mu \le -0.53$ and
$-0.96 \le \mu \le -0.89$ --- 
shows three minima at finite $\pm \lambda$ and
$\lambda=0$. 
This indicates a region of coexistence of the isotropic and nematic phases:
it reflects the negative quartic term in
the Ginzburg-Landau free energy.\cite{chung}
Hence, the transition 
between the isotropic and nematic phases is first order.  
The electron density as a function of $\mu$ shows
a discontinuity at these critical values of $\mu_c$,
which is a signature of the first order transition.

%Since 
%the electron density is fixed, the chemical potential shows a 
%discontinuity at these critical values of $x$ as well.

To better understand the behavior of the nematic order, as well as
the robustness of our result, we considered the dependence of $\lambda$
on the parameters $F_2$ and $t^{\prime}$.
Upon increasing $|F_2|$, we found the maximum value of
$\lambda$ to increase.  Also, increasing $|F_2|$ shifts 
the critical values of $\mu$, making the region over which
nematic order exists wider.  Furthermore, we found an
absence of nematic order for all doping concentrations
if $ |F_2|/(2 t) \le 6\times 10^{-3}$ (within the numerical
precision) for $t^{\prime} =-0.4 t$.
By varying $t^{\prime}$, we found the structure shown
in Fig.~\ref{fig:lambdaplot} to shift.  In particular,
when $t^{\prime} = 0$, $\lambda$ is maximal at $\mu=0$  
({\it i.e.} half-filling) and is a symmetric function
of $\mu$.  This is a consequence of the particle-hole
symmetry present when $t^{\prime} = 0$.
Though the critical values of $\mu$ and the maximal value 
of $\lambda$ depend on the value of $F_2$ and $t^{\prime}$, 
we would like to stress that the qualitative features of 
the phase transition between the isotropic and nematic 
liquids are not sensitive to the choice of parameters. 
In what follows, we will present results for 
$F_2/(2 t) =-0.55$, and $t^{\prime} =-0.4 t$ ($2t \equiv 1$).
  For small hole concentrations, this
gives a hole-like Fermi surface centered about $(\pi, \pi)$
(see Inset (a) of Fig.~\ref{fig:nhallplot}).  Such a Fermi
surface is consistent with angle resolved photoemission
spectroscopy\cite{campuzano} and Hall constant measurements
\cite{hwang} in the high-$T_c$ cuprates.

%%%%%%%%%%%%%%%%%%%%%%%%%%%%%%%%%%%%%%%%%%%%%%%%%%%%%%%%%%%%%%%%%%%%%%%%%%%%
%%%%%%%%%%%%%%%%%%%%%%%%%%%%%%%%%%%%%%%%%%%%%%%%%%%%%%%%%%%%%%%%%%%%%%%%%%%%

\section{Conductivity Tensor and the Hall Constant}

\begin{figure}
\scalebox{.51}{\includegraphics{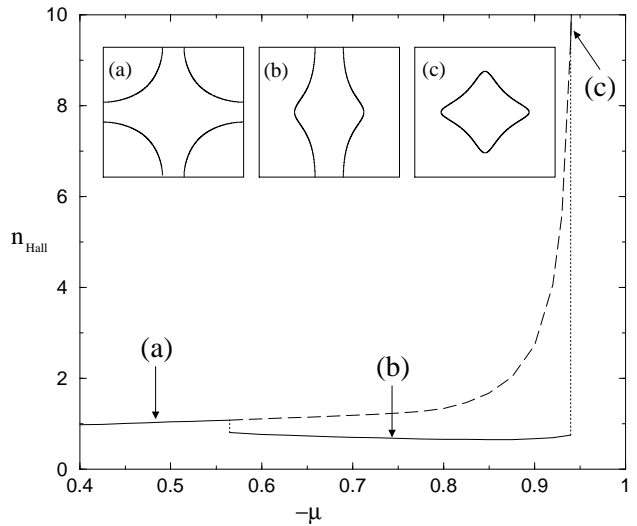} }
% \scalebox{.51}{\includegraphics{nhallplot3.eps} }
\caption{Hall constant as a function of 
chemical potential, $\mu$
for $F_2/( 2t) =-0.55$ and $t^{\prime} = -0.4 t$ ($2 t \equiv  1$).
%\cite{note}.  
The dashed line is the Hall constant
without the nematic order. The inset shows the the Fermi surface
for the points labeled by arrows in the main figure.}
\label{fig:nhallplot}
\end{figure}

Above, we saw that the nematic phase occurs in a finite
regime dopings.  Furthermore, we found a dramatic change
in the Fermi surface topology upon developing nematic 
order.  (See the Inset of Fig.~\ref{fig:nhallplot}.)  
In this section, we study the 
conductivity tensor to probe the nematic order, as
the conductivity tensor, especially the hall 
conductivity is sensitive to the topology of the 
Fermi surface.
%Since the conductivity tensor, and especially the Hall 
%conductivity, is sensitive to the topology of the Fermi 
%surface, we study change of the conductivity as the nematic phase
%develops.

Within a Boltzmann equation approach, the longitudinal 
conductivities are (in units where $\hbar = 1$)
\begin{equation}
\sigma_{ii} =  2e^2  \frac{1}{N} \sum_{\bf k}
   \left(- \frac{\partial f}{\partial \xi_{\bf k} } \right)
    \left( v^2_i \tau_{\bf k} \right),
\label{boltz}
\end{equation}
where $i$ = $x$ or $y$, $e$ is the electron's charge, $v_i$ 
is the component of the Fermi velocity along the $i$-axis,
and $\tau_{\bf k}$ is the transport lifetime. 
In the presence of a magnetic field ${\bf H}$ parallel to the 
$\hat z$-axis, the Hall conductivity is\cite{ziman,ong}
\begin{equation}
\sigma_{xy} = 2e^3 H \frac{1}{N} \sum_{\bf k}
   \left(- \frac{\partial f}{\partial \xi_{\bf k} } \right)
   \left( v_y \tau_{\bf k} \right)
   \left( {\bf v}_{\bf k} \times \nabla_{\bf k} \right)_{\hat z}
   \left( v_x \tau_{\bf k} \right)  \, .
\label{hall}
\end{equation}
In what follows, we will report results for the case
$\tau_{\bf k} \equiv \tau = {\rm const}$.  Some discussion 
regarding the form of $\tau_{\bf k}$ is given in the following
Section.

\begin{figure}
\scalebox{.75}{\includegraphics{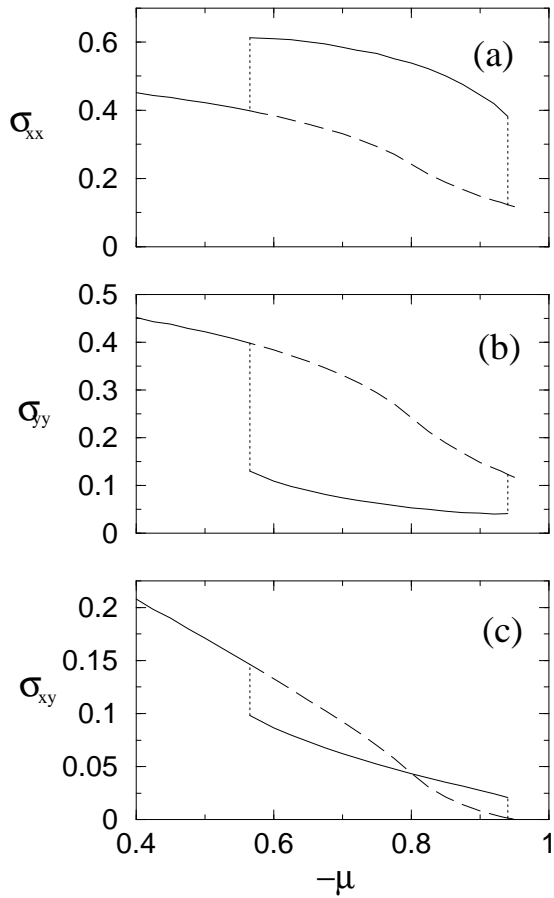} }
% \scalebox{.75}{\includegraphics{sigmaijplot1.eps} }
\caption{
The conductivity tensor as a
function of chemical potential $\mu$ for $|F_2|/(2t)
=0.55$ and $t^{\prime}=-0.4t$ ($2t \equiv  1$).
(a) $\sigma_{xx}$ vs. $\mu$, (b) $\sigma_{yy}$ vs. $\mu$,
and (c) the Hall conductivity $\sigma_{xy}$ vs. $\mu$.
$\sigma_{xx}$ and $\sigma_{yy}$ are in units of $2 e^2 (2t)^2$;
$\sigma_{xy}$ is in units of $2 e^3 H (2 t)^3$.
%The conductivities (a) $\sigma_{xx}$ along $x$-direction
%(b) $\sigma_{yy}$ along $y$-direction (in unit of $2 e^2 (2t)^2 $), 
%and (c) the Hall conductivity
%$\sigma_{xy}$  (in unit of $2 e^3 H (2 t)^3$)
%as a function of chemical potential, $\mu$
%for $|F_2|/(2 t) =0.55$ and $t^{\prime} =-0.4 t$ ($2t \equiv 1$). 
The dramatic change is due to the phase transition to nematic state. 
The dashed lines
indicate the behavior of conductivities without the nematic order. }
\label{fig:sigmaplot}
\end{figure}

Using the single-particle dispersion in Eq.~\ref{band} with 
$\lambda$ determined from Eq.~\ref{nematic}, we plot the 
conductivity tensor vs. $\mu$ in Fig.~\ref{fig:sigmaplot}.  
For comparison, the conductivity tensor without nematic
order ($\lambda = 0$) is shown as a dashed line.  
When nematic order develops, the Fermi surface topology
changes dramatically.  More specifically, the hole-like
Fermi surface with nematic order becomes an open Fermi 
surface upon developing nematic order.  
In the nematic state,
the Fermi surface is rather flat along the $\hat y$-direction.
(See Inset (b) of Fig.~\ref{fig:nhallplot}.)  Hence, the
hole's motion along the $\hat x$-direction is enhanced,
while it is suppressed along the $\hat y$-direction.
These changes in the Fermi surface are reflected in the 
longitudinal conductivities, Figs.~\ref{fig:sigmaplot}(a)
and \ref{fig:sigmaplot}(b) --- $\sigma_{xx}$ increases
suddenly with the onset of nematic order (at $\mu=-0.57$),
while $\sigma_{yy}$ shows a sudden decrease.
The Hall conductivity, Fig.~\ref{fig:sigmaplot}(c), shows
a dramatic drop at $\mu=-0.57$ as well.  However, its 
change is small compared with $\sigma_{xx}$ and $\sigma_{yy}$.
This occurs because the Hall conductivity is not only 
sensitive to the Fermi surface's topology, but also its
curvature.\cite{ong}  As shown by the dashed line in 
Fig.~\ref{fig:sigmaplot}(c), the Hall conductivity already
decreases fairly rapidly without nematic order.
The dotted lines at the transition indicate that
Boltzmann equation, Eq. ~\ref{boltz} is not applicable
near the transition, since there are two phases with
different densities coexist.

% We will cast the Hall constant as a function of single variable
% of doping concentration, $x$, to make a specific connection to
% high $T_c$ cuprates, where the nematic Fermi liquid was 
% suggested.\cite{kivelson2}  

Another quantity of interest is the Hall constant $n_H$, 
which measures the number of charge carriers, and is related 
to the Hall resistivity.  For weak magnetic fields, the Hall 
resistivity is
\begin{equation}
\rho_{xy} \simeq \frac{\sigma_{xy}}{\sigma_{xx}\sigma_{yy}}
= \frac{1}{n_Hec} H,
\end{equation}
which defines $n_H$ in terms of $\rho_{xy}$.
For the case of a single-band Fermi liquid model with a
circular (or elliptical) Fermi surface, $n_H$ is the 
density of charge carriers per unit cell {\it i.e.} the 
area of the Fermi surface.

In Fig.~\ref{fig:nhallplot}, we plot the Hall constant, $n_H$ 
as a function of chemical potential, $\mu$.
For comparison, the Hall number without nematic
order ($\lambda = 0$) is shown as a dashed line.  
Notice the sharp drop of $n_H$ at $\mu=-0.57$, where nematic order
becomes finite.   
It is expected that $n_H$ is almost constant in the
region where nematic order exists, $-0.94 \leq \mu \leq -0.57$.
This can be understood from the conductivity tensor in 
Fig.~\ref{fig:sigmaplot} --- the decrease of hole motion 
along the $y$ direction is compensated by the increase of
hole motion along $x$ direction, giving a product 
$\sigma_{xx}\sigma_{yy}/\sigma_{xy} \sim n_H$ which is 
nearly constant.

% As the nematicity, $\lambda$ increases, the Fermi surface gets 
% distorted and it will eventually open up at the zone boundary 
% at a critical $\lambda_c$ which depends on the doping concentration.
% We found that the Fermi surface in the nematic phase for 
% $ 0.17 \le x \le 0.6$ is a open Fermi surface as shown in the 
% inset, (b) of Fig. 3.  We expect that the transport properties 
% will reflect the existence of nematic order.

%%%%%%%%%%%%%%%%%%%%%%%%%%%%%%%%%%%%%%%%%%%%%%%%%%%%%%%%%%%%%%%%%%%%%%%%%%%%
%%%%%%%%%%%%%%%%%%%%%%%%%%%%%%%%%%%%%%%%%%%%%%%%%%%%%%%%%%%%%%%%%%%%%%%%%%%%

\section{Discussion}

To begin with, our results for the phase transition in 
Fig.~\ref{fig:lambdaplot} were obtained within a mean-field 
approximation.  Once fluctuations about the mean-field state 
are included, we expect the range over which the nematic phase 
exists to be reduced.  Furthermore, fluctuations will reduce 
the value of $\lambda$. The effect of fluctuations on the nature
of phase transition will be addressed in near future.
Another approximation in our treatment was the neglect of
collective modes.  In the continuum limit of Eq. \ref{model},
the breaking of rotational symmetry leads to a Goldstone mode;
scattering from this Goldstone mode gives rise to non-Fermi 
liquid behavior.\cite{vadim}  On a lattice, however, this 
collective mode is gapped; we expect scattering from the 
collective mode to have little effect on the $dc$ conductivities.  
%However, the effect of this mode on other physical quantities 
%at finite frequency could be important, and is an interesting 
%problem for future study.
%
% While the nature of the phase transition is rather robust to 
% the choice of the parameters, $|F_2|/t$ and $t^{\prime}/t$, 
% the qualitative feature of the Hall constant can be altered 
% by changing the $t^{\prime}/t$.\cite{note} 

%%%%%%%%%

The subject of disorder deserves some discussion.  
Though disorder was introduced to understand the behavior of
the conductivity through the transitions, 
we did not address its effects on the transition between 
the isotropic and nematic phases, nor did we address its 
consequences on the nematic phase itself.\cite{note2}
In the absence of disorder, the mean-field treatment of 
the Hamiltonian Eq. \ref{model} gave the transition between 
the isotropic and nematic phases to be first order.  
For systems that undergo a first order transition without
disorder, it was
shown that the system with quenched disorder could be
mapped onto a random field Ising model.\cite{cardy}
Based on this mapping, it was shown that
in  $d \le 2$ dimensions, 
the coexistence at the phase transition disappears 
for arbitrarily small amount of disorder;
in  $d > 2$ dimensions 
this occurs by the application of a finite amount of disorder.
%The nature of the phase transition/crossover in the presence
%of the quenched disorder is still needed to be investigated.
%
For the transition in this work, however,
there are gapless fermions presents on both sides of the transition.
Therefore it is unclear to us how much the results obtained from
the mapping may apply.
We hope to address these issues in the future.
%
%(for discrete microscopic degrees of freedom), 
%coexistence at the phase transition disappears, 
%and the phase transition is converted from first order
%to second order, 
%by the application of a finite amount of disorder,
%\cite{berker}
%Therefore, 
%in $d > 2$ dimensions, the disorder will eliminate the discontinuities
%by the application of a finite amount of bond randomness, and
%in $d \le 2$, this occurs with  arbitrarily small amounts of quenched
%disorder.
%However, it is not clear whether the quenched disorder will 
%convert the first order transition to the second order transition,
%or it will turn the phase transition to the crossover for $d=2$.\cite{cardy}
%Moreover, the effect of the quenched disorder on quantum phase transition
%is a little explored area, and it is a subject for
%future study.

%
%However, there are still many open questions regarding the effect of
%quenched disorder on quantum phase transitions, and  whose pure versions are
%first order transitions, when disorder couples to a local
%energy density. 
%
%While understanding the nature of the transition with quenched randomness
%for both $d=2$ and $d> 2$ is a subject of further study,
%it is likely that 
%the significant drop in the conductivity tensor and $n_H$ will be weakened
%in the presence of the quenched disorder,
%
%The signature of the nematic order in Hall constant will 
%remain as a dip in the Hall constant in a much narrow range 
%due to the fluctuation and disorder.\cite{note}

It is also important to note that the conductivity tensor 
depends of the form of the transport scattering rate, 
and hence on the disorder.  Realistic models for disorder 
usually give transport scattering rates which are momentum 
dependent.  In particular, in most high-$T_c$ materials, 
disorder is mainly due to the dopant ions, which lie outside 
of the $CuO_2$ planes.  As a result, these dopant ions cause 
primarily forward scattering.
In the above discussion, we described our results for the
case $\tau_{\bf k} \equiv \tau$.  However, we have also 
considered the case where $\tau_{\bf k} \sim |{\bf v}_{\bf k}|$.
\cite{chung}  Such a form arises for out-of-plane disorder, 
causing primarily forward scattering.\cite{kee2}
When $\lambda = 0$, we found the overall features of $n_H$ 
to be very different from $\tau_{\bf k} =\tau$.  However, 
the effect of the nematic order on the Hall constant gives 
the same result as when $\tau_{\bf k} \equiv \tau$ --- 
there is a sharp drop in the Hall constant in the nematic 
phase compared with the isotropic phase.  

%%%%%%%%%%%%%%%%

Our findings for the Hall constant have potential relevance 
to the experiments of the normal state Hall resistivity in 
the high-$T_c$ superconductor, 
Bi$_2$Sr$_{2-x}$La$_x$CuO$_{6+\delta}$.\cite{ando,sudip}
In these experiments, the Hall constant was measured in the
normal state at low temperatures, by suppressing superconductivity 
with a large pulsed magnetic field of 60 T.
They observed saturation of the Hall constant at low temperatures,
suggesting conventional transport properties.   
However, the Hall constant as a function of doping concentration shows
a maximum at $x=0.15$, and  decreased sharply in underdoped regime.  
%These results provide
%evidence for a phase transition in the normal state of the 
%high-$T_c$ cuprates.  The dramatic drop of the Hall constant in 
%the underdoped regime was emphasized, and a possible interpretation 
%was given in Ref.~\onlinecite{sudip}.  However, it should be noted  
In overdoped regions,
there is a regime near optimal doping $x \simeq 0.16$, where 
the Hall constant shows the dip --- the decrease is about $20 \%$ ($\simeq 0.2$) of its 
maximum value ($\simeq 1.0$) at $x=0.15$.  The dip of the Hall 
constant around $x=0.16$ becomes more pronounced at low temperatures 
($ < 1 K$).  
This suggests that the nematic phase may be a candidate
to explain the anomaly observed in the Hall constant in 
high $T_c$ cuprates.

%%%%%%%%%%%

In conclusion, to study the nematic phase we considered
the Hamiltonian in Eq.~\ref{model} with quadrupolar 
density-density interactions on a square lattice.
Within a mean-field analysis, we found the nematic phase
to occur in a finite region of dopings.  Moreover, the 
transition between the isotropic and nematic phases was
found to be first order; a dramatic change in the Fermi
surface topology accompanied this transition.
As the conductivity is sensitive to the topology of the
Fermi surface, we investigated the effects of nematic
order on transport.  
We found the conductivity tensor and Hall constant to 
show significant changes at the transitions between the
isotropic and nematic phases.  If observed, such 
signatures would give evidence for nematic order;
it would give evidence for the approach pursued in this
paper as an appropriate description of nematic phases
in physical systems.  
Finally, we discussed the effect of disorder and the
scattering from collective modes on the phase transitions
and on the transport properties at the transitions.

%%%%%%%%%%%%%%%%%%%%%%%%%%%%%%%%%%%%%%%%%%%%%%%%%%%%%%%%%%%%%%%%%%%%%%%%%%%%
%%%%%%%%%%%%%%%%%%%%%%%%%%%%%%%%%%%%%%%%%%%%%%%%%%%%%%%%%%%%%%%%%%%%%%%%%%%%

{\it Acknowledgments}
We would like to thank S. A. Kivelson, V. Oganesyan
P. Coleman,  Y-B Kim, and E. Fradkin for insightful discussions, 
and Yoichi Ando for useful comments. 
We also thank P. Fournier and W. Metzner for pointing out 
Ref.~\onlinecite{hwang} and Ref. ~\onlinecite{metzner}, respectively.
This work was supported by the NSERC of Canada(HYK,EHK,CHC),
Canada Research Chair(HYK), Canadian Institute of Advanced Research(HYK),
Alfred P. Sloan Research Fellowship (HYK), and 
Emerging Material Knowledge program funded by Materials and Manufacturing
Ontario (HYK).

%%%%%%%%%%%%%%%%%%%%%%%%%%%%%%%%%%%%%%%%%%%%%%%%%%%%%%%%%%%%%%%%%%%%%%%%%%%%
%%%%%%%%%%%%%%%%%%%%%%%%%%%%%%%%%%%%%%%%%%%%%%%%%%%%%%%%%%%%%%%%%%%%%%%%%%%%

%%%%%%%%%%%%%%%%%%%%%%%%%%%%%%%%%%%%%%%%%%%%%%%%%%%%%%%%%%%%%%%%%%%%%%%%%%%%
%%%%%%%%%%%%%%%%%%%%%%%%%%%%%%%%%%%%%%%%%%%%%%%%%%%%%%%%%%%%%%%%%%%%%%%%%%%%

\end{document}